\documentclass[prb,twocolumn,superscriptaddress,showpacs,amsmath,amssymb]{revtex4-2}

\usepackage{graphicx}
\usepackage{latexsym}
\usepackage{amsmath}
\usepackage{amssymb}
\usepackage{amsfonts}
\usepackage{color}
\usepackage{bm}
\usepackage{verbatim}
\usepackage{float}

\definecolor{MS-color}{RGB}{255,0,0}

\usepackage{framed}
\definecolor{shadecolor}{RGB}{222,222,221}
\begin{document}
 
  
\title{Anderson-Higgs mass of magnons  in superconductor/ferromagnet/superconductor systems  }

\date{\today}

 \author{Mikhail Silaev}
\affiliation{Computational Physics Laboratory, Physics Unit, Faculty of Engineering and Natural Sciences, Tampere University, P.O. Box 692, Tampere, Finland}

 \begin{abstract}
Anderson-Higgs mechanism of mass generation is a generic concept in high-energy and condensed matter physics. 
  It shows up through the Meissner effect providing the expulsion of static and low-frequency 
  magnetic fields from superconductors. However,  it does not affect propagating 
  electromagnetic waves with a  spectrum gap determined by the plasma frequency, which is too large to be sensitive to the 
  superconducting transition.   
   Here we demonstrate the spectroscopic manifestation of the Anderson-Higgs mass, showing that it 
   determines the spectrum gap of magnons in superconductor/ferromagnet/superconductor multilayers. 
   Moreover, we show that this effect has been observed in recent experiments as a spontaneous ferromagnetic resonance frequency shift in such systems.
 Our theory explains many unusual experimental features and suggests 
  effective controls over the magnon spectrum with tunable spectral gap and group-velocity reversal.
   These findings pave the way to a wide range of advanced functionalities for possible applications in magnonics. 
   \end{abstract}

 \maketitle

 \section{Introduction}
  
 The concept of spontaneous symmetry breaking and mass generation in local gauge theories plays the central roles both in the particle physics and in condensed matter systems. 
  The Anderson-Higgs (AH) mechanism \citep{altland2010condensed} of mass generation  resulting from the spontaneous breaking of local gauge symmetries takes place both in 
charged superconductors \cite{anderson1958random, anderson1963plasmons} and  in the  Standard Model 
\cite{englert1964broken, higgs1964broken,weinberg1967model}. 
However there is  an  important  conceptual difference between these cases.
 In the high-energy theories the vector boson masses are determined solely by the AH mechanism. On the contrary, in metals the propagating photons have an energy gap  equal to the  plasma frequency already in the normal state in result of interactions\cite{ anderson1963plasmons}. It is determined by the total density of electrons and therefore is not sensitive to the superconducting transition. 
 Rather, superconductivity shows up in the  Meissner effect, that is the  modification of the photon decay length at frequencies much smaller than the plasma one. 
  Therefore when it comes to the  condensed matter systems it is fair to say the AH mass of propagating excitations  has never been  observed.

In this Letter we demonstrate that the energy gap due to the AH mechanism can be directly observed for magnons in hybrid systems consisting of coupled  metallic ferromagnet (F) and superconducting (S) layers. The electromagnetic interaction between the dynamical magnetization  and the superconducting condensate leads to the spontaneous generation of magnon spectral gap in the absence of any magnetic anisotropy. We argue that this phenomenon has been detected in recent experiments as the spontaneous shift of the ferromagnetic resonance (FMR) frequency controlled by the superconducting order parameter in S/F/S systems, with Nb and Py materials as S and F, respectively\cite{li2018possible,jeon2019effect,golovchanskiy2020magnetization} . 
 Thus the superconductivity can be used for the manipulation of magnon states for spin wave logic \cite{chumak2021roadmap} and magnonics\cite{barman20212021,chumak2022roadmap,
pirro2021advances}. Our proposal provides a  step towards applications of superconducting spintronics \cite{bergeret2005odd, Buzdin2005,Linder2015,beckmann2016spin, quay2018out,ohnishi2020spin,han2019spin,RevModPhys.90.041001},  superconducting magnonics\cite{bespalov2014magnon,jeon2018enhanced,ojajarvi2022dynamics,johnsen2021magnon,prokopenko2019recent, dobrovolskiy2019magnon, dobrovolskiy2021cherenkov, golovchanskiy2020magnetization,
 jeon2019effect,li2018possible, muller2021temperature,golovchanskiy2018ferromagnet,
golovchanskiy2019ferromagnet} and and on-chip cavity magnonics \cite{golovchanskiy2021approaching,golovchanskiy2021ultrastrong,
li2019strong,hou2019strong, rameshti2022cavity} applications. 
  
 \begin{figure}[htb!]
  $\begin{array}{c}
\includegraphics[width=1.0\linewidth]
 {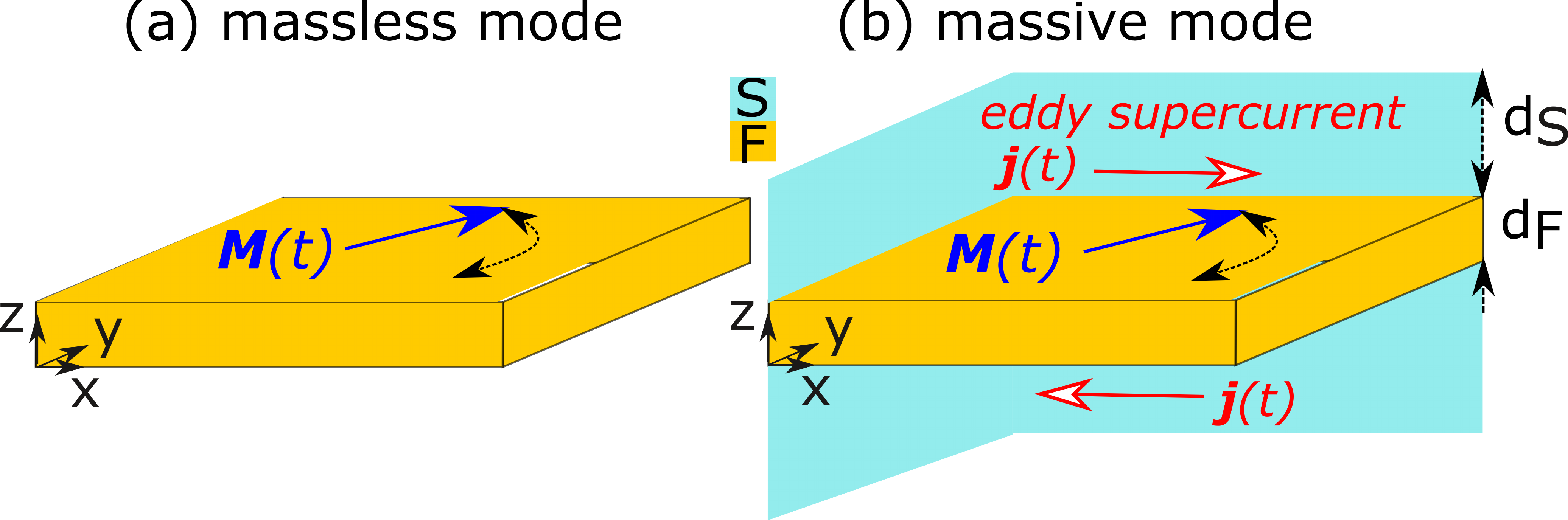}  
 \end{array}
 $
 \caption{\label{Fig:1} 
 (a) In the F slab the in-plane magnetization rotation $\bm M (t)$ i is a zero-mass mode.
 (b)  In the S/F/S system  the in-plane magnetization rotation $\bm M (t)$
 is massive mode because of the the eddy supercurrents $\bm j (t)$. }
 \end{figure}

\section{Qualitative picture} 
\label{Sec:Qualitative}

We start with  the qualitative argument. 
    Consider the F slab with in-plane magnetization  in the absence of external field and any in-plane anisotropies shown in Fig.\ref{Fig:1}a. The  magnetization rotation in $xy$ plane $\bm M = M_0 (\cos\theta, \sin\theta, 0)$ is a massless mode 
  which means that the  magnetic energy is degenerate by the direction $\theta$, which plays the role of the U(1) gauge field. In result, magnon spectrum  is also massless having no energy gap. Consider e.g. magnetostatic surface spin waves in ferromagnet slab in the 
  Damon-Eshbach geometry\cite{damon1961magnetostatic}  when the  external magnetic  field $\bm H_0$ is aligned in $xy$ plane $\bm H_0\perp \bm z$ and perpendicular to the
magnon wave vector $\bm q \perp \bm H_0$. Magnon dispersion is given by
 \cite{prabhakar2009spin}
 \begin{align} \label{Eq:magnons}
  \Omega_{mag} (q)= \sqrt{ \Omega_K^2 + 
  \frac{\Omega_M^2}{4}(1- e^{-2qd_F})  } 
 \end{align}
 where $\Omega_K = \gamma \sqrt{H_0 (H_0 + 4\pi M_0)}$ is the Kittel frequency\cite{kittel1948theory} of the FMR in F slab,
 $\Omega_M = 4\pi \gamma M_0$ and $M_0$ is the magnetization of F. 
 For vanishing external field $H_0 \to 0$ the minimal frequency vanishes $\Omega_{mag}(q=0) \to 0$ corresponding to the massless mode. 
  
 The situation changes in the presence of superconductors in contact with  ferromagnetic film as shown in Fig.\ref{Fig:1}b. 
   According to the Faraday law  
time-dependent magnetization components $\bm M (t) = \bm M_0 + \bm M_\omega e^{i\omega t}$ generate the  
  curly electric field outside the F film 
 $\bm E_\omega  = \pm (2\pi i\omega/c) d_F \bm z \times \bm M_\omega $
 with upper sign for $z>d_F/2$ and lower sign for $z<-d_F/2$.   This electric field  induces eddy currents in the adjacent metallic layers\cite{kennewell2007calculation,kostylev2009strong}. In the present case of S layers these are the eddy supercurrents $\bm j_\omega = \sigma_S(\omega) \bm E_\omega$ where the  frequency-dependent conductivity of S is given by \cite{TinkhamBook}
 \begin{align} \label{Eq:sigmaS}
 \sigma_S (\omega) = \frac{ c^2}{4\pi i \lambda^2 \omega} .
 \end{align}  
 Here $\lambda$ is the London penetration length. We assume the frequency to be much smaller than the superconducting gap so that the normal component contribution can be neglected. 
 In result eddy supercurrents are given by $\bm j_\omega = \pm (cd_F/\lambda^2) \bm M_\omega\times\bm z$. Remarkably, the prefactor coefficient here does not depend on frequency. Thus, assuming for simplicity the thickness of S layers to me small $d_S\ll \lambda$ we get the contribution of the kinetic energy of the eddy supercurrents contributing to the 
 action describing the magnetic degrees of freedom normalized by the F thickness
  \begin{align} \label{Eq:Action}
   S_m =  - 2\pi M_0^2\frac{d_S d_F}{\lambda^2}\int d t d\bm r \theta^2
  \end{align}
  where the angle $\theta$ determines the deviation of  $\bm M (t)$ from its stationary direction. 
 Eq.(\ref{Eq:Action}) is the AH mass-generating  term for magnons. 
 
  The Goldstone mode which is "absorbed" by the U(1) gauge field $\theta$ to get Eq.\ref{Eq:Action} is the Josephson phase difference 
  $\varphi $ between the upper and lower S layers in Fig.\ref{Fig:1}b. Because of the strong exchange field and large F layer thickness the  Josephson energy is negligible.  In general 
   the action Eq.(\ref{Eq:Action}) has the gauge-invariant form obtained by replacing $\theta \to  \theta + L_m\nabla_x\varphi$ where $L_m= \phi_0/(4\pi^2d_F M_0)$ and $\phi_0$ is the flux quantum.
 The time-dependent Josephson phase difference $\varphi (t)$ generates the voltage 
 upper and lower S layers $ V (t) = (\phi_0/2\pi) \partial_t \varphi $ 
 between upper and lower S layers.  
 In case of the metallic F with conductivity $\sigma_F$ due to the skin effect this ac voltage $V$ and hence $\varphi$ 
 are localized   near the edges of S/F/S structure at the skin length ${\rm Im} l_{sk}$ where  $l_{sk} = c\sqrt{1/4\pi i \omega \sigma_F}$. 
 Away from this shell $V=0$ and $\varphi=0$ so that the action has the form Eq.(\ref{Eq:Action}). 
 

\section{Model}
 
To put the above qualitative reasoning on the solid ground
we consider below the exact calculation of the FMR fundamental mode and magnetostatic waves       
 in S/F/S structures. 
 Our starting equations for the frequency components
 of magnetization $\bm M_\omega$, current $\bm j_\omega$, magnetic 
 $\bm H_\omega$ and electric $\bm E_\omega$ fields
\begin{align} \label{Eq:LLG}
 {\rm LLG:\;\;\;} & i\omega {\bm M}_\omega = \gamma(\bm B_0\times\bm M_\omega + \bm B_\omega \times\bm M_0  )
 \\
  \label{Eq:HomegaGen}
 {\rm in\; metal:\;\;\;}  
 & \bm\nabla\times (\sigma^{-1}\bm\nabla\times 
 \bm H_\omega ) +  \frac{4\pi i \omega}{c^2} 
 \bm B_\omega =  0
 \\ \label{Eq:HomegaVac}
 {\rm in\; insulator:\;\;\;} 
 & \bm\nabla\times \bm H_\omega =0
 \end{align} 
This system consists of the Landau-Lifshitz-Gilbert (LLG) Eq.\ref{Eq:LLG}, where $\bm B_0 = \bm H_0 + 4\pi \bm M_0$ is the stationary magnetic field.  Length scales are assumed larger than the exchange length and gradient terms in LLG are neglected. 
Maxwell equations in metal taken in the usual  quasi-static approximation\cite{landau2013electrodynamics} yield Eq.\ref{Eq:HomegaGen}. 
The conductivity is $\sigma_F= const $ in F while in S it
  is given by Eq.(\ref{Eq:sigmaS}). 
   In the vacuum or insulator where $\sigma= 0$ we neglect the displacement current as well and use the Maxwell  Eq.\ref{Eq:HomegaVac} which yields that  $\bm H_\omega $ is a potential field. This approximation is valid provided the lateral size along $x$ is much smaller than the wavelength at the given frequency. 
   For $\omega \sim 10 $GHz it is $c/\omega \sim 1$ cm. 
 
 \begin{figure}[htb!]
 \centerline{
 $ \begin{array}{c}
 \includegraphics[width=1.0\linewidth]
 {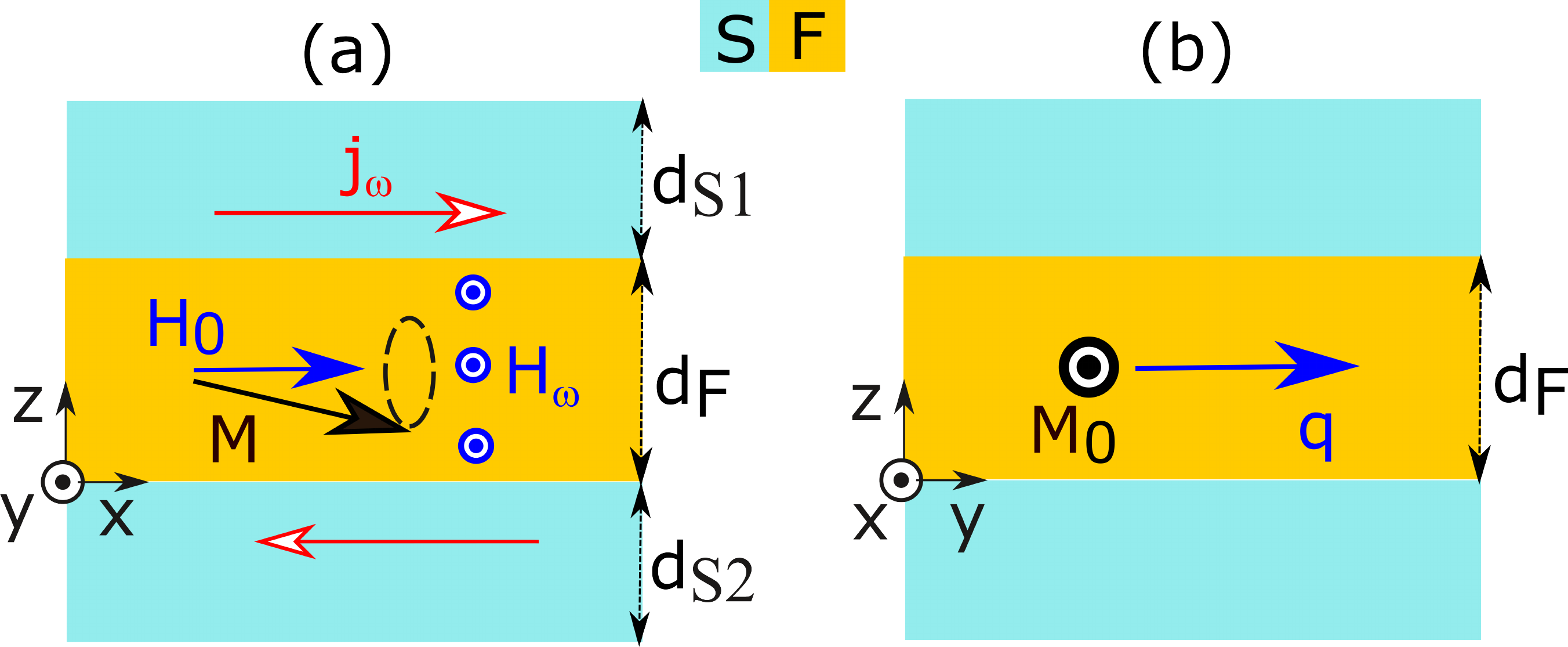}
 \end{array}$
  }
  \caption{\label{Fig:FMRSFS} 
 (a) Ferromagnetic resonance in S/F/S system. 
 Distributions of  loop eddy currents $\bm j_\omega$ and the magnetic field that they generate $\bm H_\omega$ is shown schematically.
 (b) The magnetostatic mode in S/F/S structure in Danone-Eschbach geometry with wave vector $\bm q\perp \bm M_0$. }
    \end{figure}
 
 \begin{figure*}[htb!]
 \centerline{
 $ \begin{array}{c}
 \includegraphics[width=0.8\linewidth] {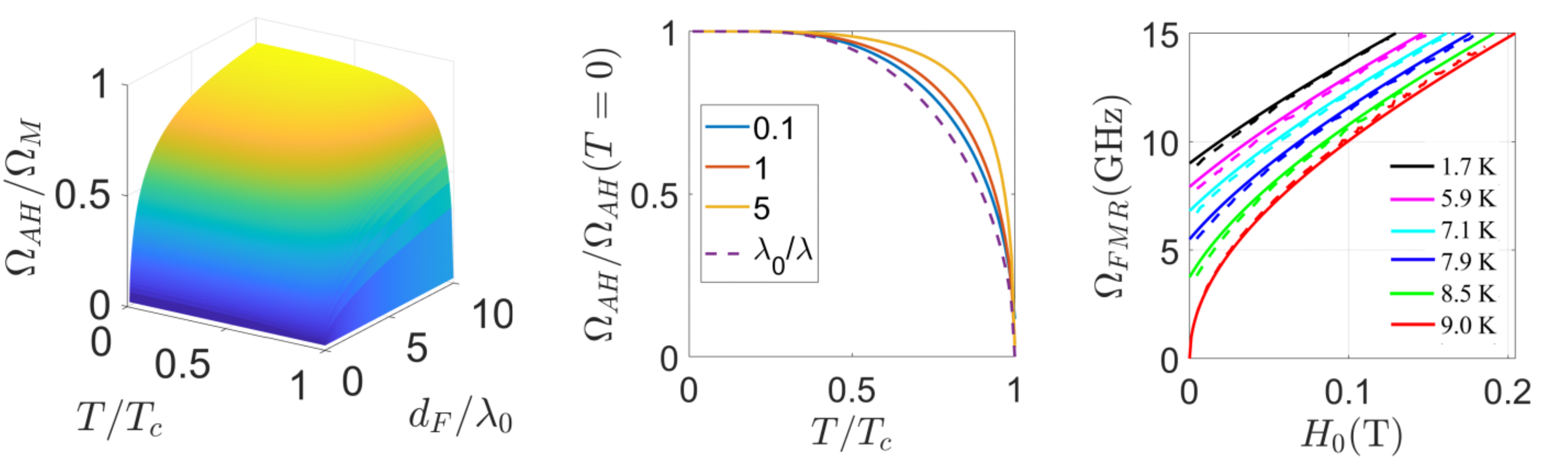}
  \put (-350,120) {\large (a)}
 \put (-200,120) {\large (b)}
   \put (-60,120) {\large (c)}
  \end{array}$
 }
 \caption{\label{Fig:ResultsFMRSFS}
(a,b) Anderson-Higgs mass of magnons $\Omega_{AH}$ given by Eq.\ref{Eq:AHmass}.
  (a) $\Omega_{AH} (T,d_F)$  normalized by $\Omega_M = 4\pi \gamma M_0$, showing saturation at  $T\ll T_c$ and 
 $d_F\gg \lambda$. 
 (b) $\Omega_{AH} (T)$ for $d_F/\lambda_0 = 0.1;\; 1;\; 5$ shown by solid lines compared with the $\lambda_0/\lambda (T)$ shown by dashed line, where $\lambda_0=\lambda (T=0)$.  
 (c) Comparison of $\Omega_{FMR}(H_0)$ given by Eq.(\ref{Eq:GeneralFMRsfs}) shown by solid lines with experimental data from Fig.2c of
 Ref.\cite{golovchanskiy2020magnetization} shown by dashed lines. Parameters are detailed in the text. 
  }
 \end{figure*}
 
 \section{Fundamental FMR mode}
 The fundamental FMR mode is then given by the eigen mode of the system Eq.(\ref{Eq:HomegaGen}) where all fields depend only in $z$. 
 We assume the lateral sizes of S/F/S structure to be sufficiently large to neglect edge effects.  
 Time-dependent magnetization components are $\bm M_\omega = (0, M_y, M_z)$. The $B_z$ component is continuous so that $B_z=0$ and therefore $H_z = - 4\pi M_z$ everywhere. 
 Under such conditions solving the LLG Eq.\ref{Eq:LLG} in terms of $H_y$ yields $B_y= (\Omega_B^2 -\omega^2)/(\Omega_K^2 -\omega^2) H_y$, 
 where $\Omega_B = \gamma B_0$. 
 The  field $H_y \neq 0$ is characteristic for metallic ferromagnets \cite{kennewell2007calculation} and provides the detuning of FMR frequency from the Kittel one $\Omega_K$.  Substituting the found $B_y$ into the Eq.\ref{Eq:HomegaGen} we get 
 \begin{align} 
  \label{Eq:HyF}
 {\rm in\; F:}\;\;\;\;\;\; & \nabla_z^2 H_y - l_{F}^{-2} H_y=0
 \\ \label{Eq:HyS}
 {\rm in\; S:}\;\;\;\;\;\; & \nabla^2_z  H_y - \lambda^{-2}  H_y = 0
  \end{align}  
   Here the renormalized skin depth length in F is determined by
 \begin{align} \label{Eq:SkinDepthFerro}
 {l}_F= l_{sk} \sqrt{ (\Omega_K^2 -\omega^2)/(\Omega_B^2 - \omega^2)}
 \end{align} 
   Boundary conditions at F/S interfaces follow from Eq.(\ref{Eq:HomegaGen})
   \begin{align} \label{Eq:BCHy1}
   & H_y |_F = H_y |_S
   \\ \label{Eq:BCHy2}
   & \nabla_z H_y |_F = \frac{\sigma_F}{\sigma_S} \nabla_z H_y |_S
   \end{align}
   Boundary conditions at the S/I and F/I interfaces, where I stands for the insulating layer or vacuum is $H_y|_I=0$. 
   Indeed, in the considered case when all fields do not depend on $y$  we obtain from Eq.(\ref{Eq:HomegaVac}) 
   $\bm \nabla\times \bm H_\omega= \bm z \nabla_x H_y - \bm x \nabla_z H_y - \bm y \nabla_x H_z =0  $
   so that $\nabla_x H_y =0 $. Integrating this equation along $x$ from any point in the I layer to the infinity outside the S/F/S structure 
   where $H_y(x=\infty)=0$  we get that  $H_y =0$ everywhere in I region. 
   The eigen mode of the system Eqs.(\ref{Eq:HyF},\ref{Eq:HyS}) with boundary conditions determine the FMR frequency.  

We start with the simplest symmetric case when  $d_{S1}=d_{S2}=d_S$ in Fig.\ref{Fig:FMRSFS}a. 
The solution of Eqs.(\ref{Eq:HyF}, \ref{Eq:HyS}) is 
symmetric about the middle of F  $z=0$
\begin{align}
  & \text{in F:}\;\;\;\;  H_y (z) = \tilde H_F \cosh (z/ l_F)  
  \\
  & \text{in S:}\;\;\;\; H_y (z) = \tilde H_S \sinh [( z-d_S - d_F/2)/\lambda] .
\end{align}
Matching solutions at $z=d_F/2$ using Eqs.(\ref{Eq:BCHy1},\ref{Eq:BCHy2})
and using the identity $\sigma_S\lambda^2/\sigma_F l_{sk}^2 =1$ we get the general equation determining 
the FMR frequency 
 \begin{align} \label{Eq:GeneralSymmetricSFS}
 \tanh ( d_S/\lambda) \tanh (d_F/2l_F) = - \lambda l_F/l_{sk}^2
  \end{align}

 For frequencies $\omega< 100$ GHz and conductivity\cite{mayadas1974resistivity} of Py $\sigma_F \sim 10^6 $ Ohm$^{-1}$m$^{-1}$ the non-magnetic metal skin depth is $l_{sk} > 10\; \mu$m. Thus for realistic F layer thickness $d_F < 1\;\mu$m the condition  $d_F\ll l_{sk}$ is satisfied. If the frequency is far from the Kittel mode we have $l_F\sim l_{sk}$ and based on the above estimations can expand 
$\tanh (d_F/2l_F) \approx d_F/2l_F$. 
Thus we obtain from Eq.(\ref{Eq:GeneralSymmetricSFS}) the explicit expression for the FMR frequency 
  \begin{align} \label{Eq:GeneralFMRsfs}
    &  \Omega_{FMR}= \gamma\sqrt{ \frac{
   B_0^2 (d_F/2\lambda) \tanh ( d_S/\lambda)  +  B_0H_0}
    {(d_F/2\lambda) \tanh ( d_S/\lambda)  +1} } 
    \end{align}
  The general Eq.\ref{Eq:GeneralFMRsfs} is qualitatively different from the Kittel expression\cite{kittel1948theory} with generalized demagnetizing factors \cite{mironov2021giant}.  
  For small external field $H_0 \ll B_0$ and F layer thickens $d_F\ll \lambda\tanh(d_S/\lambda)$ Eq.\ref{Eq:GeneralFMRsfs} reduces to the Kittel expression\cite{kittel1948theory}
with the artificial anisotropy energy $E_a=K_a M_y^2$,
where
  $K_a=2\pi (d_F/\lambda)\tanh (d_S/\lambda )$. 
  In case $d_S\ll \lambda$ we get $E_a=2\pi  M_0^2 (d_Fd_S/\lambda^2) \theta^2 $ in agreement with Eq.\ref{Eq:Action}.  
In general the AH mass of magnons $\Omega_{AH}$ is given by the FMR frequency shift
 $\Omega_{AH}=\Omega_{FMR} (H_0=0)$ so that
 \begin{align} \label{Eq:AHmass}
      \Omega_{AH}= 4\pi  \gamma M_0 \sqrt{ \frac{
    (d_F/2\lambda) \tanh ( d_S/\lambda)  }
    {(d_F/2\lambda) \tanh ( d_S/\lambda)  +1} }
 \end{align}

  At larger F thickness $d_F\gg \lambda\coth(d_S/\lambda)$
 the AH magnon mass  Eq.(\ref{Eq:GeneralFMRsfs}) saturates
 at $\Omega_{AH} = 4\pi \gamma M_0$. Recently the tendency to such saturation has been observed experimentally \cite{golovchanskiy2022magnetization} in Nb/Py/Nb multilayers.
 Taking into account the temperature dependence of $\lambda (T)$ 
  calculated within the model of diffusive superconductor we get the behaviour $\Omega_{AH} (d_F, T)$ shown in
   Fig.\ref{Fig:ResultsFMRSFS}a,b. The fixed   parameters relevant for the Nb/Py/Nb experimental systems\cite{li2018possible,jeon2019effect,golovchanskiy2020magnetization} are $d_S = 150$ nm, $\lambda (T=0) = 80$ nm,  
   $4\pi M_0 = 1.06$ T.   From Fig.\ref{Fig:ResultsFMRSFS}b it is clear that the temperature dependencies of $\Omega (T)$ are rather close to that of the inverse London length in $\lambda^{-1} (T)$ 
  in accordance with experiments \cite{golovchanskiy2020magnetization,golovchanskiy2022magnetization}. 
   
  As it is clear from Eq.(\ref{Eq:AHmass}) and Fig.\ref{Fig:ResultsFMRSFS}a the scale of $\Omega_{AH}$ is determined  by
   $\Omega_M= 4\pi \gamma M_0$ which for Py is $\Omega_M \approx 30$ GHz. Thus, the observed FMR frequency 
   shifts\cite{golovchanskiy2020magnetization, golovchanskiy2022magnetization} of the order
   $10$ GHz are explained by theory. To be more convincing, in Fig.\ref{Fig:ResultsFMRSFS}c
   we show the direct numerical comparison of theoretical $\Omega_{FMR} (H_0)$
   dependencies given by Eq.(\ref{Eq:GeneralFMRsfs}) with experimental data presented in Fig.2c of Ref.\cite{golovchanskiy2020magnetization}.
     Theoretical curves and experimental data are shown by solid and 
   dashed lines, respectively. 
  As one can see, the present theory provides very accurate fits of the experimental $\Omega_{FMR} (H_0)$
  dependencies for all temperatures. All parameters except $d_S$  are chosen exactly the same 
  as in the experiment\cite{golovchanskiy2020magnetization} $T_c = 9$ K, 
    $\lambda (T=0) = 80$ nm,  $d_F=19$ nm,   $4\pi M_0 = 1.06$ T so that $\Omega_M = 31.3$ GHz. 
  The only parameter we adjust to get most accurate fits in Fig.\ref{Fig:ResultsFMRSFS}c  is   
   $d_S = 80$ nm which is a bit smaller than in the experiment $d_S = 110$ nm. 
   This is natural given there is a superconductivity suppression  near the S/F interfaces at the scale of
    coherence length \cite{kittel1996introduction} which is  $\xi_0 \sim  10-40$ nm in Nb.
      
Now let us consider a more general case when S layers thickness are different $d_{S1} \neq d_{S2}$. A bit more involved calculation \cite{supplement} yields 
 \begin{align} \label{Eq:GeneralFMRsfsd1d2}
 &  \Omega_{FMR}= 
 \\ \nonumber
 & \gamma\sqrt{ \frac{
 B_0^2 (d_F/\lambda) [ \coth ( d_{S1}/\lambda) 
 + 
 \coth ( d_{S2}/\lambda)]^{-1}   +  B_0H_0}
 {(d_F/\lambda) [ \coth ( d_{S1}/\lambda) + 
 \coth ( d_{S2}/\lambda)]^{-1}  +1} } 
 \end{align}  
 This expression describes how FMR shift disappears together with the thickness of one of the S layers with the transition from S/F/S to S/F structure. If e.g. $d_{S1}\to 0$ Eq.\ref{Eq:GeneralFMRsfsd1d2} yields the usual Kittel frequency $\Omega_{FMR} \to \Omega_K =\gamma \sqrt{H_0B_0}$. Thus, the FMR frequency of S/F structure is equal to $\Omega_K$. 

In the considered approximation of the structure length along $x$ being much shorter than the wavelength the FMR mode  in S/F 
is equivalent to that in S/F/I/S structure, where I is the insulator. This is because in both cases there is the same boundary condition $H_y=0$ at the F/I or F/vacuum interface.  This conclusion explains experiments \cite{li2018possible,jeon2019effect,golovchanskiy2020magnetization} where seemingly enigmatic suppression $\Omega_{AH}$ by the I layer has been observed. This phenomenon has an electromagnetic origin not connected with the influence of spin-triplet proximity effect induced by the magnetic precession\cite{houzet2008ferromagnetic, li2018possible}. 
Similarly, there is no effect of superconductivity on the FMR mode in S/FI/S structures where FI stands for the  ferromagnetic insulator, e.g. quite popular material YIG\cite{serga2010yig}. In this case $H_y=0$ everywhere so that no currents are generated in the adjacent S layers. However, superconductivity affects the dispersion of spin waves in such setups \cite{yu2022efficient}.

 \section{Magnetostatic magnons in S/F/S system}
  
Now let us turn to the spin waves in S/F/S system in the geometry 
Fig.\ref{Fig:FMRSFS}b with $\bm M_0 \parallel \bm x $ searching the solution in the form where all fields $\propto e^{ iq_y y }$. 
In this case the calculation becomes more involved as compared to FMR due to the presence of both $B_y$ and $B_z$ components. 
From Eq.\ref{Eq:HomegaGen}, see the detailed derivation\cite{supplement} we get
 \begin{align} 
  \label{Eq:HyFmag}
 {\rm in\; F:}\;\;\;\;\;\; & \nabla_z^2 H_y - (l_{F}^{-2} +q^2) H_y=0
 \\ \label{Eq:HySmag}
 {\rm in\; S:}\;\;\;\;\;\; & \nabla^2_z  H_y - (\lambda^{-2}+q^2)  H_y = 0 ,
  \end{align}  
  supplemented by the boundary conditions at S/F interfaces which consist of 
 continuity  Eq.\ref{Eq:BCHy1} and 
  \begin{align}  \label{Eq:BCHy2mag}
  & (\nabla_z H_y - iqH_z) |_F = 
  \frac{\sigma_F}{\sigma_S} (\nabla_z H_y - iqH_z) |_S
  \end{align}

We consider the range of momenta $q l_{sk} \gg 1$ which is experimentally relevant given that $l_{sk} \sim 10 \mu$m. For simplicity we assume  $ d_{S1,2}\gg \lambda$. In this approximation it is possible to find magnon dispersion solving Eqs.(\ref{Eq:HyFmag}, \ref{Eq:HySmag}, 
\ref{Eq:BCHy2mag}) analytically\cite{supplement}
\begin{align} \label{Eq:MagnonGen}
 & \Omega_{mag} (q) = 
  \sqrt{ \frac{ (\Omega_M + a\Omega_0 )^2 - (\Omega_M + b\Omega_0 )^2 e^{-2qd_F} }
 { a^2 -b^2 e^{-2qd_F}  } } 
\end{align}
where $\Omega_0 =\gamma H_0 $ and $a = 1+  q \lambda /\sqrt{1 +q^2 \lambda^2 }$, 
$b= 1- q \lambda /\sqrt{1 +(q \lambda)^2 }$.
 Expression Eq.(\ref{Eq:MagnonGen}) yields two limiting cases. First, at large momenta $q\lambda \gg 1$ it reduces to the usual surface spin wave dispersion Eq.(\ref{Eq:magnons}). The reason is that the fields outside F decay at scales
 $q^{-1}\ll \lambda$ and therefore cannot excite eddy supercurrents.  
  Second, at $q=0$
 the magnon frequency Eq.(\ref{Eq:MagnonGen}) coincides with the FMR frequency Eq.(\ref{Eq:GeneralFMRsfs}). 
 In the absence of the external field $H_0=0$ it yields the  AH magnon mass
 given by Eq.(\ref{Eq:AHmass}) $\Omega_{mag} (q=0, H_0=0) =\Omega_{AH}$ taken  in the limit $ d_S \gg \lambda$. 
 
 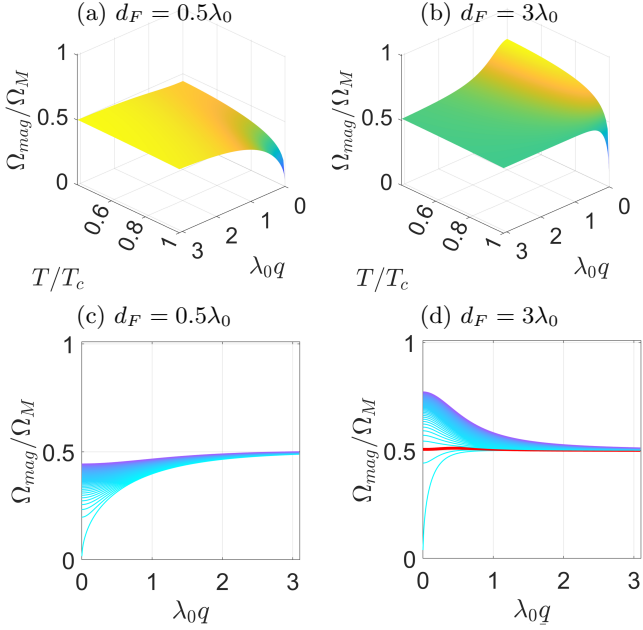
\begin{figure}[htb!]
 \centerline{
 $ \begin{array}{c}
 \includegraphics[width=1.0\linewidth] {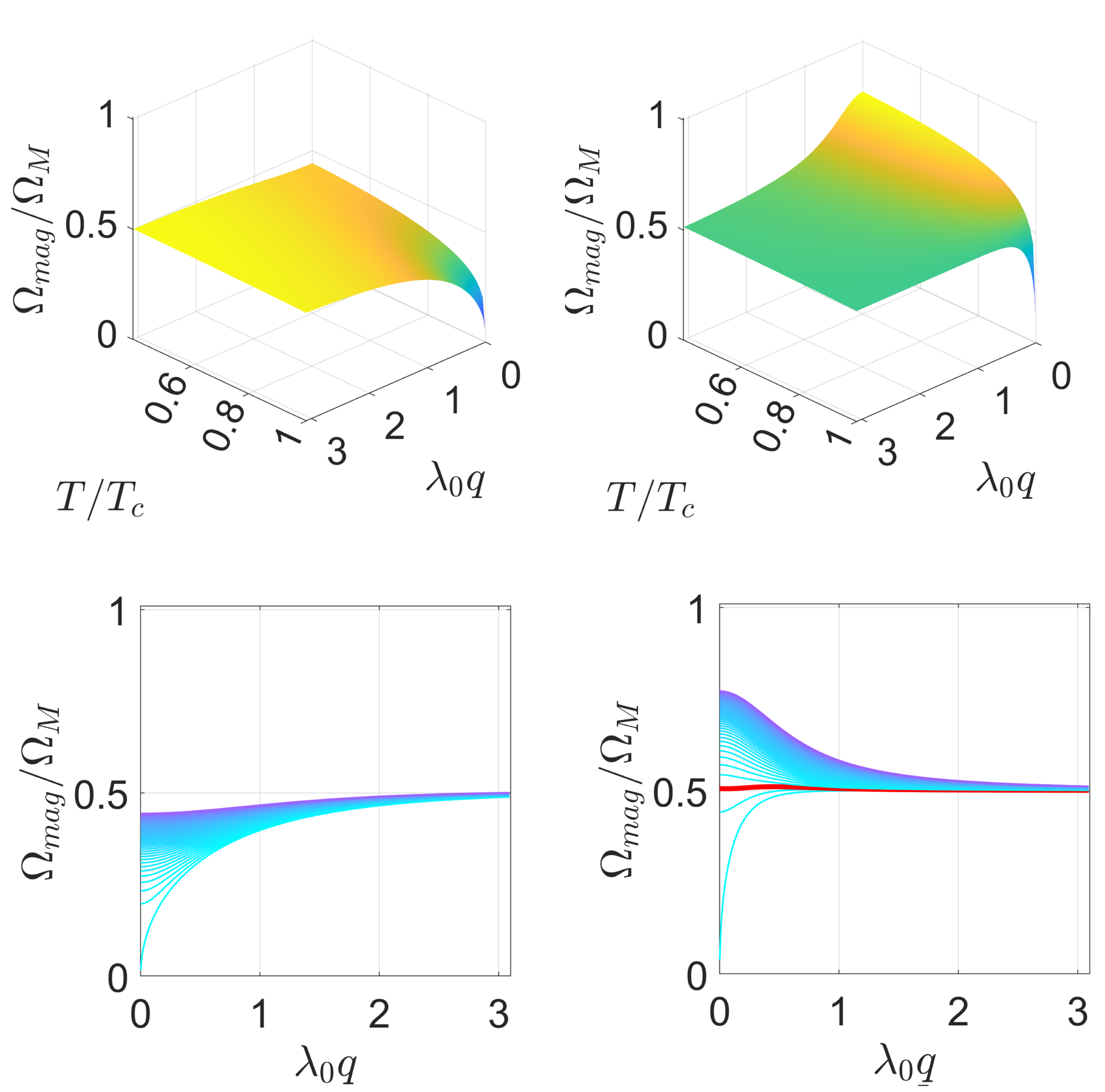}
 \put (-220,230) {  (a) $d_F=0.5 \lambda_0$}
 \put (-90,230) { (b) $d_F=3 \lambda_0$}
 \put (-220,115) {  (c) $d_F=0.5 \lambda_0$}
 \put (-90,115) {  (d) $d_F=3 \lambda_0$}
 \end{array}$
 }
 \caption{\label{Fig:MagnonDisp} 
 The evolution of dispersion $\Omega_{mag} (q)$ of surface spin waves in S/F/S system at different temperatures at $H_0=0$. (a) $d_F=0.5\lambda_0$ and (b) $d_F=3\lambda_0$. In the latter case the dispersion change from "forward"  $v_{g}>0$ to "backward" $v_g<0$ with decreasing temperature. (c,d) The set of curves $\Omega_{mag} (q)$ for different temperatures starting from $T=T_c$ (lower curve) to $T=0.5 T_c$ (upper curve)
  with the step $0.01 T_c$. The almost flat magnon spectrum shown by red thick curve in (d) corresponds to the threshold at $T=0.98 T_c$. 
   }
 \end{figure}
 
 Comparing $\Omega_{mag}(q=0) = \Omega_{AH}$  and the  asymptotic   $\Omega_{mag}(q\to \infty) = \Omega_{M}/2$ we note that there are two qualitatively different cases. For $\Omega_{AH}<\Omega_{M}/2 $ the dependence $\Omega_{mag}(q)$ is monotonically increasing which is usual for
 surface spin waves   \citep{damon1961magnetostatic, kalinikos1990dipole, karenowska2015magnon}. 
 Such waves are "forward" ones since they have parallel group velocity and wave vector $v_{g} = \partial \Omega_{mag}/\partial q >0$. 
 On the other hand, for $\Omega_{AH}>\Omega_{M}/2 $ the dependence $\Omega_{mag}(q)$ 
first becomes non-monotonic and then at even larger $\Omega_{AH}$ it is monotonically decreasing. This regime corresponds to the "backward" wave  $v_{g} <0$ with antiparallel group velocity and wave vector. Provided that $d_F$ is sufficiently large this transition in magnon spectrum occurs at a certain threshold temperature determined by 
 $\Omega_{AH} (T) = \Omega_{M}/2$. 
 Shown on Fig.\ref{Fig:MagnonDisp} are the two characteristic cases 
 of (a) $d_F= 0.5 \lambda_0$ when  waves are "forward" for all temperatures and (b) $d_F= 3 \lambda_0$ when there is a crossover from "forward" to "backward" regime at $T\approx .98 T_c$. At this temperature there is almost flat magnon dispersion as shown in Fig.\ref{Fig:MagnonDisp}d by the red thick curve.   

  
  \section{Summary}
To summarize, we demonstrate that there exists an energy gap of magnons in S/F/S systems, which appears spontaneously in the absence of 
 external magnetic field or magnetic anisotropy.
This magnon gap is interpreted as AH mass since it emerges upon the gauge-symmetry breaking phase transition into the superconducting state. Previously the AH mechanism in condensed matter systems has been commonly associated with the 
 Meissner effect responsible for the expulsion of electromagnetic fields from superconductors at frequencies 
 substantially smaller than the plasma frequency threshold for propagating waves in metals.  
 Our results show that AH mass can also be observed in a qualitatively different way by {\it spectroscopic} means through the modification 
 of propagating magnon dispersion. 
Moreover,  I show that AH mass has already been observed in a series of recent experiments \cite{li2018possible,jeon2019effect, zhao2020exploring, golovchanskiy2020magnetization, golovchanskiy2022magnetization} on FMR frequency shift in Nb/Py/Nb structures. 
The non-zero FMR frequency at $H_0=0$  has been detected and remained enigmatic until now. The present work establishes that this frequency shift is precisely the AH mass of magnons and explains all the unusual features observed experimentally. The derived 
Eq.(\ref{Eq:GeneralFMRsfs}) provides accurate fits of experimental data \cite{golovchanskiy2020magnetization}.

Proposed controls of spin waves with the help of superconductivity pave the way to various functionalities. 
For example, it is possible to localize magnons and fabricate magnon crystals by patterning one of the S layers. Alternatively, one can use light to modulate the order parameter and the AH magnon mass. It will enable the creation of optically-controlled magnon resonators and crystals. Moving this resonator along the F film and implementing magnons' transport will be possible by shifting the focused light spot. Besides these examples, one

\section{Acknowledgements} 
Many stimulating discussions with Igor Golovchanskiy, Vladmir Krasnov and Alexander Mel'nikov were very useful for this work . 
 
 \appendix
 
 \section{FMR in case of different S layer thickness}
  \label{Sec:Appendix1}
  
 We use the same equations and boundary conditions as in the main text. 
 In S layers the solution is  
 \begin{align}
  & \text{in S1:}\;\;\;\; H_y (z) = \tilde H_{S1} \sinh 
  ((z-d_{S1}-d_F/2)/\lambda) 
  \\
  & \text{in S2:}\;\;\;\; H_y (z) = \tilde H_{S1} 
  \sinh ((z+d_{S2}+d_F/2)/\lambda) 
\end{align}

   From the boundary condition at S/F interface we get for the field in F
 \begin{align}
 ( \partial_z H_y /H_y )(x=d_F/2) =-\frac{\sigma_F}{\lambda\sigma_S} \coth (d_{S1}/\lambda)  
  \\ 
  ( \partial_z H_y /H_y )(x=-d_F/2) =\frac{\sigma_F}{\lambda\sigma_S} \coth (d_{S2}/\lambda) 
 \end{align}
 The general solution in F is 
 \begin{align}
  H_y = A e^{z/l_F} + B e^{-z/l_F} 
 \end{align}
 From the above b.c. we get 
 \begin{align}
 & \frac{l_{sk}^2}{\lambda l_F} \frac{A e^{d_F/2l_F} + B e^{- d_F/2l_F} }
 {A e^{d_F/2l_F} - B e^{- d_F/2l_F}} =  -\tanh (d_{S1}/ \lambda)
 \\
  & \frac{l_{sk}^2}{\lambda l_F} \frac{A e^{-d_F/2l_F} + B e^{ d_F/2l_F} }
 {A e^{-d_F/2l_F} - B e^{d_F/2l_F}} = \tanh (d_{S2}/ \lambda)
 \end{align}
 Solving with respect to the coefficients $A,B$ we get 
 \begin{align}
  A e^{d_F/2l_{F}} [ 1  + (l_{sk}^2/\lambda l_F) \tanh (d_{S1}/ \lambda) ] = 
  \\  \nonumber
  - B e^{-d_F/2l_{F}} [ 1  - (l_{sk}^2/\lambda l_F) \tanh (d_{S1}/ \lambda) ] 
  \\ 
   A e^{-d_F/2l_{F}} [ 1  - (l_{sk}^2/\lambda l_F) \tanh (d_{S2}/ \lambda) ] = 
  \\ \nonumber
    - B e^{d_F/2l_{F}} [ 1  + (l_{sk}^2/\lambda l_F) \tanh (d_{S2}/ \lambda) ] 
 \end{align}  
  Dividing one of these equations by another we get rid of the coefficients and remain with the equation 
  \begin{align} \label{AppEq:FMRd1d2Gen}
  & e^{2d_F/l_{F}} = 
  \frac{1- (l_{sk}^2/\lambda l_F) ( a_1+a_2 ) + 
  (l_{sk}^2/\lambda l_F)^2 a_1a_2   }
   {1+ (l_{sk}^2/\lambda l_F) ( a_1+a_2 )+ 
   (l_{sk}^2/\lambda l_F)^2 a_1a_2  }
  \end{align}
  where $a_1 = \tanh (d_{S1}/ \lambda)$ and $a_2 = \tanh (d_{S2}/ \lambda)$.
  Further we assume that  $l_{sk}^2\gg \lambda l_F$ which means there should be detuning from Kittel frequency larger than $\Omega_M \lambda/l_{sk} \sim 10^{-3} \Omega_M$. 
  Then from Eq. \ref{AppEq:FMRd1d2Gen} we obtain 
  \begin{align}
   \frac{\tanh (d_{S1}/ \lambda) \tanh (d_{S2}/ \lambda)}{\tanh (d_{S1}/ \lambda) + \tanh (d_{S2}/ \lambda)} = - \frac{\lambda}{d_F} \frac{l_F^2}{l_{sk}^2}
  \end{align} 
  Substituting $l_F$ we get 
  \begin{align}
   \frac{\omega^2 -\Omega_B^2}{\omega^2 -\Omega_K^2} = - 
   \frac{d_F}{\lambda}[ \coth (d_{S1}/\lambda) + \coth (d_{S2}/\lambda) ] 
    \end{align}
 Solving this equation with respect to $\omega$ we get Eq.17 in the main text.

     \section{Spin waves} 
     \subsection{Basic equations} 
     \label{AppSec:SpinWavesBasicEqs}
     
     First, the Maxwell equation in components yields in both S and F 
  \begin{align} \label{AppEq:MaxwellMagComponentsY}
  & \nabla_z ( iq  H_z - \nabla_z H_y ) = -\frac{4\pi \sigma}{c^2} i\omega B_y 
  \\\label{AppEq:MaxwellMagComponentsZ}
  & iq ( iq H_z - \nabla_z H_y ) = \frac{4\pi \sigma}{c^2} i\omega B_z
\end{align}      
 with boundary conditions 
 \begin{align}\label{AppEq:MagBC}
  ( iq  H_z - \nabla_z H_y )|_F = \frac{\sigma_F}{\sigma_S}( iq  H_z - \nabla_z H_y )|_S 
 \end{align}
 Now we can transform  Eqs.(\ref{AppEq:MaxwellMagComponentsZ},\ref{AppEq:MaxwellMagComponentsY}) 
   in S and F regions separately. 
  
  {\bf  First,} in S $\bm B=\bm H$ 
   so that  
   \begin{align} \label{AppEq:MaxwellMagComponentsYS}
  &  iq \nabla_z H_z - \nabla^2_z H_y  = -\lambda^{-2} H_y 
  \\
  \label{AppEq:MaxwellMagComponentsZS}
  & q^2 H_z + iq \nabla_z H_y  = -\lambda^{-2} H_z
  \end{align}      
  From the condition $\bm \nabla \cdot \bm H =0$ we get $\nabla_z H_z =- iq H_y $ and transform the equations further 
   \begin{align} \label{AppEq:MaxwellMagSHy}
   \nabla^2_z H_y = \lambda^{-2}_q H_y
\end{align}     
where $\lambda^{-2}_q= q^2 + \lambda^{-2} $. 
Further we assume that the thickness of S layer is larger than $\lambda_g$ so that only the decaying mode exist there 
   $\partial_z H_z = \mp \lambda^{-1}_q H_z$ so that $H_z = \pm i\lambda_q q H_y$. Then the r.h.s. of the b.c. \ref{AppEq:MagBC} is 
   \begin{align}\label{AppEq:BC}
 iq H_z - \nabla_z H_y = \pm  \lambda_q \lambda^{-2} H_y 
   \end{align}
   where the upper and lower signs correspond to $z=\pm d_F/2$. 
    
 {  \bf  Second,} in F Maxwell equations are 
    \begin{align} \label{AppEq:MaxwellMagComponentsYF}
  &  iq \nabla_z H_z - \nabla^2_z H_y  = -l^{-2}_{sk} B_y 
  \\
  \label{AppEq:MaxwellMagComponentsZF}
  & q^2 H_z + iq \nabla_z H_y  = -l^{-2}_{sk} B_z
  \end{align}     
  From LLG equations we get 
   \begin{align}
  & B_y= \frac{H_y (\Omega_K^2 - \omega^2) +  i \omega \Omega_M H_z}   
  {\Omega_0^2- \omega^2} 
   \\
    & B_z= \frac{H_z (\Omega_K^2 - \omega^2) -  i \omega \Omega_M H_y}
    {\Omega_0^2- \omega^2} 
\end{align}    
   Then Maxwell equations are written as 
   \begin{align} \label{AppEq:MaxwellHy}
 & -\left( l_{sk}^2q^2 + 
  \frac{\omega^2-\Omega_K^2}{\omega^2-\Omega_0^2} \right) H_z 
  = 
  il_{sk}^2 q  \nabla_z H_y + 
  \frac{ i\omega \Omega_M}{\omega^2 - \Omega_0^2}  H_y 
  \\ \label{AppEq:MaxwellHz}
 & \left( l_{sk}^2\nabla_z^2 -  
  \frac{\omega^2-\Omega_K^2}{\omega^2-\Omega_0^2} \right) H_y 
  = \left( q l_{sk}^2 \nabla_z 
  - 
  \frac{ \omega \Omega_M}{\omega^2 - \Omega_0^2} \right) i H_z 
  \end{align}
  From Eq.\ref{AppEq:MaxwellHy} we express $H_z$
  \begin{align} \label{AppEq:Hz}
  & i H_z =  \frac{q(\omega^2 - \Omega_0^2) \nabla_z H_y 
  + 
  \omega l^{-2}_{sk}  \Omega_M H_y}
  {(\omega^2 - \Omega_0^2)q^2  + l_{sk}^{-2} (\omega^2-\Omega_K^2)} 
 \end{align}    
 Substituting this equation to (\ref{AppEq:MaxwellHz}) we get in the r.h.s.
 \begin{align}
 & \left( l_{sk}^2 q \nabla_z  - 
  \frac{ \omega \Omega_M}{\omega^2 - \Omega_0^2}  \right) i H_z  = 
  \\ \nonumber
 & \frac{l_{sk}^2q^2(\omega^2 - \Omega_0^2) \nabla_z^2 H_y + 
  q \omega \Omega_M \nabla_z H_y}
  {(\omega^2 - \Omega_0^2)q^2  + l_{sk}^{-2} (\omega^2-\Omega_K^2)} 
  -
  \\ \nonumber
 &  \frac{ \omega \Omega_M}{\omega^2 - \Omega_0^2} 
   \frac{q(\omega^2 - \Omega_0^2) \nabla_z H_y 
   + 
  \omega l^{-2}_{sk}  \Omega_M H_y}
  {(\omega^2 - \Omega_0^2)q^2  + l_{sk}^{-2} (\omega^2-\Omega_K^2)} = 
  \\ \nonumber
 & \frac{
 l_{sk}^2q^2(\omega^2 - \Omega_0^2)^2 \nabla_z^2 H_y 
 -
 (\omega \Omega_M)^2 l^{-2}_{sk} H_y
  }
  {(\omega^2 - \Omega_0^2)[(\omega^2 - \Omega_0^2)q^2  + l_{sk}^{-2} (\omega^2-\Omega_K^2)]}
 \end{align}
  Further we get 
  \begin{align}
   \nabla_z^2 H_y = \left[ q^2 +  l_{sk}^{-2} 
   \frac{(\omega-\Omega_K)^2 - ( \omega \Omega_M)^2 }
   {(\omega^2-\Omega_K^2)(\omega^2-\Omega_0^2)} \right] H_y
  \end{align}
   The last term can be transformed as 
   \begin{align}
  & (\omega^2-\Omega_K^2)^2 - ( \omega\Omega_M)^2 = 
   \\ \nonumber
  & (\omega^2 - \Omega_0^2)^2 - 8\pi (\omega^2 - \Omega_0^2)\Omega_M\Omega_0 
    - (\Omega_M)^2 ( \omega^2 - \Omega_0^2)  =
    \\ \nonumber
    & (\omega^2-\Omega_0^2)(\omega^2-\Omega_B^2) 
   \end{align}
  Hence, Maxwell equation finally is 
   \begin{align}\label{AppEq:MaxwellF}
   \nabla_z^2 H_y = \left( q^2 +  l_{F}^{-2}  \right) H_y
  \end{align}
  which is the Eq.18 in the main text.
  
  Further, we use Eq.\ref{AppEq:Hz}
 to transform the boundary condition Eq.\ref{AppEq:BC}
  \begin{align}
  ( iq H_z - \partial_z H_y)|_F = 
  \frac{(\Omega_K^2-\omega^2)\nabla_z H_y +  \omega \Omega_M H_y}{(\omega^2-\Omega_e^2) q^2l_{sk}^2 + (\omega^2-\Omega_K^2) } 
  \end{align} 
  so that 
  \begin{align}
  \frac{(\Omega_K^2-\omega^2)\nabla_z H_y +  \omega \Omega_M H_y}{(\omega^2-\Omega_e^2) q^2l_{sk}^2 + (\omega^2-\Omega_K^2) } = 
  \pm \frac{\lambda_q}{l_{sk}^2} H_y
  \end{align}
   This boundary condition is simplified in the limit $ql_{sk} \gg 1$
\begin{align} \label{AppEq:MagBC1}
  \frac{\Omega_K^2-\omega^2}{\omega^2-\Omega_e^2} \nabla_z H_y 
  + 
  \frac{\omega \Omega_M}{\omega^2-\Omega_e^2} H_y
  = 
  \pm \lambda_q q^2 H_y
  \end{align}
  
  \subsection{Derivation of dispersion relation}
   \label{AppSec:SpinWavesDerivation}  
  
   To derive the dispersion of spin waves we solve the system 
   (\ref{AppEq:MaxwellF}, \ref{AppEq:MagBC}). In the limit 
   $q l_{sk} \gg 1$ we search the solutions in F as 
 \begin{align}
 H_y = A e^{qz} + B e^{-qz}
 \end{align}
 The b.c. yields 
 \begin{align} \label{AppEq:ABsystemEqS1}
 & (a + \chi + \kappa ) e^{qd} A  = (b + \chi - \kappa ) e^{-qd} B 
 \\ \label{AppEq:ABsystemEqS2}
 & (b + \chi + \kappa ) e^{-qd} A  = (a + \chi - \kappa ) e^{qd} B    
 \end{align}
   which yields the equation 
   \begin{align} \label{AppEq:Dispersion0}
   (a  + \chi )^2 -\kappa^2 = 
   [(b  + \chi )^2 -\kappa^2] e^{-2qd_F}
   \end{align}
   where we denote $a=1+q\lambda_q$, $b=1-q\lambda_q$, 
   $\chi =  \Omega_0 \Omega_M /(\Omega_0^2- \omega^2)$ 
   and $\kappa  =  \omega \Omega_M /(\Omega_0^2- \omega^2)$. 
   We can transform the above equation 
 \begin{align}
 & ( a  + \chi )^2 -\kappa^2  = 
 \\ \nonumber
 & a^2 +2a\frac{\Omega_0\Omega_M}{\Omega_0^2-\omega^2} + 
 \frac{(\Omega_0\Omega_M)^2}{(\Omega_0^2-\omega^2)^2} - 
 \frac{(\omega\Omega_M)^2}{(\Omega_0^2-\omega^2)^2} = 
 \\ \nonumber
 & \frac{a^2 (\Omega_0^2 - \omega^2 ) + 
 2a\Omega_0\Omega_M + \Omega_M^2}{\Omega_0^2-\omega^2} =
 \\ \nonumber
 & \frac{( a \Omega_0 + \Omega_M )^2 -a^2\omega^2}{\Omega_0^2-\omega^2} 
 \end{align}
 Hence the Eq.\ref{AppEq:Dispersion0} can be written as 
 \begin{align}
 \frac{a^2\omega^2 - (a\Omega_0 + \Omega_M)^2 }{b^2\omega^2 - 
 (b\Omega_0 + \Omega_M)^2} = e^{-2q d_F}
 \end{align}
 which yields the solution 
 \begin{align} \label{AppEq:MagnonGen}
 & \Omega_{mag} (q) = 
  \sqrt{ \frac{ (\Omega_M + a\Omega_0 )^2 - 
  (\Omega_M + b\Omega_0 )^2 e^{-2qd_F} }
 { a^2 -b^2 e^{-2qd_F}  } } 
 \end{align}
 
%
%

 \bibliographystyle{apsrev4-2}
  \bibliography{refs2}

 \end{document}